\RequirePackage{lineno}
\documentclass[aps,prl,twocolumn,superscriptaddress,groupedaddress]{revtex4}  
\usepackage{graphicx}  
\usepackage{makecell}
\usepackage{bm}        
\usepackage{amssymb}   
\usepackage{multirow}
\usepackage{epsfig}
\usepackage{amsmath}
\usepackage{float}
\usepackage{color}
\usepackage[colorlinks, linkcolor=blue]{hyperref}

\newcommand{\effstw}{\ensuremath{\sin^2\theta_{\text{eff}}^{\text{$\ell$}}}}

\newcommand{\etal}{{\it et al.}}
\newcommand{\topL}{\Xhline{0.8pt}}
\newcommand{\midL}{\Xhline{0.4pt}}
\newcommand{\botL}{\Xhline{0.8pt}}
\newcommand{\LINESPREADSIZE}{1.3}

\begin{document}

\lefthyphenmin=2
\righthyphenmin=2


\title{Measurement of the effective weak mixing angle at the CEPC}
\affiliation{Department of Modern Physics, University of Science and Technology of China, Anhui, China}
\affiliation{Institution of High Energy Physics, Chinese Academy of Sciences}
\author{Zhenyu Zhao}\affiliation{Department of Modern Physics, University of Science and Technology of China, Anhui, China}
\author{Siqi Yang}\affiliation{Department of Modern Physics, University of Science and Technology of China, Anhui, China}
\author{Manqi Ruan}\affiliation{Institution of High Energy Physics, Chinese Academy of Sciences}
\author{Minghui Liu}\affiliation{Department of Modern Physics, University of Science and Technology of China, Anhui, China}
\author{Liang Han}\affiliation{Department of Modern Physics, University of Science and Technology of China, Anhui, China}

\begin{abstract}

We present a study of the measurement of the effective weak mixing angle parameter ($\effstw$) 
at the Circular Electron Positron Collider (CEPC). As a fundamental physics parameter, $\effstw$ plays 
a key role not only in the global test of the standard model electroweak sector, but also in constraining the 
potential beyond standard model new physics at high energy frontier. 
CEPC proposes a two year running period around the Z boson mass pole at high instataneous luminosity,
providing a large data sample with $4\times 10^{12}$ $Z$ candidates generated in total.
It allows a high precision measurement of $\sin^2\theta^{\ell}_\text{eff}$ both
in the lepton and quark final states, of which the uncertainty can be one order of magnitude
lower than any previous measurement at the LEP, SLC, Tevatron and LHC.
It will not only improve the overall precision of the $\sin^2\theta^\ell_\text{eff}$ experimental determination
to be comparable to the preicision of the theoretical calculation with two-loop radiative corrections,
but also provide direct comparisons between different final states.
In this paper, we also study the measurement of $\effstw$ at high mass region. With one month data 
taken, the precision of $\effstw$ measured at 130 GeV from $b$ quark final state is 0.00010, which 
will be an important experimental observation on the energy-running effect of $\effstw$.

\end{abstract}
\maketitle

\section{Introduction}

The weak mixing angle, $\theta_W$, is one of the fundamental parameters of the standard model (SM). It governs the 
relative strength of the axial-vector couplings to the vector couplings in the neutral-current interactions 
with Lagrangian
\begin{eqnarray}
  \mathcal{L} = -i \frac{g}{2\cos\theta_W}\bar{f}\gamma^\mu \left( g^f_V - g^f_A \gamma_5\right) f Z_\mu, 
\end{eqnarray} 
where $g^f_A$ and $g^f_V$ are the axial-vector and vector couplings, defined as $g^f_A=I^f_3$ and 
$g^f_V = I^f_3 - 2Q_f\cdot \sin^2\theta_W$ where $I^f_3$ and $Q_f$ are the weak isospin component 
and the charge of the fermion $f$. $Z_\mu$ describes the $Z$ boson exchange. To include higher order 
electroweak radiative corrections, the effective weak mixing angles are defined as 
\begin{eqnarray}
 \sin^2\theta^f_\text{eff} = \kappa_f \sin^2\theta_W,
\end{eqnarray}
where $\kappa_f$ is a flavor-dependent effective scaling factor absorbing the higher order corrections~\cite{LEP-SLD}. 
By doing this, $\sin^2\theta^f_\text{eff}$ can be directly measured from experimental observations, thus is very sensitive to 
both the precision validation of SM and the search for new physics beyond SM.

It is as custom to quote the leptonic effective weak mixing angle $\sin^2\theta^\ell_\text{eff}$,
so that measurements at the LEP, SLC, Tevatron and LHC can be directly compared with each other.
To do this, shifts between $\sin^2\theta^\ell_\text{eff}$ and $\sin^2\theta^q_\text{eff}$ need to be
calculated under the standard model assumptions.
In this work, it is calculated using the {\sc zfitter} package~\cite{ZFITTER},
which gives a shift of $-0.0001$ and $-0.0002$ for $\sin^2\theta^u_\text{eff}$ and $\sin^2\theta^d_\text{eff}$
with respect to $\sin^2\theta^\ell_\text{eff}$, respectively.
For $\sin^2\theta^b_\text{eff}$ as a special case, the shift is $+0.0014$E.
Such calculation has a high precision as long as the energy of the interaction is not approaching 160 GeV where
the correction from box-diagrams becomes sizable~\cite{flavorstw}.

\begin{figure}[!hbt]
\begin{center}
\epsfig{scale=0.4, file=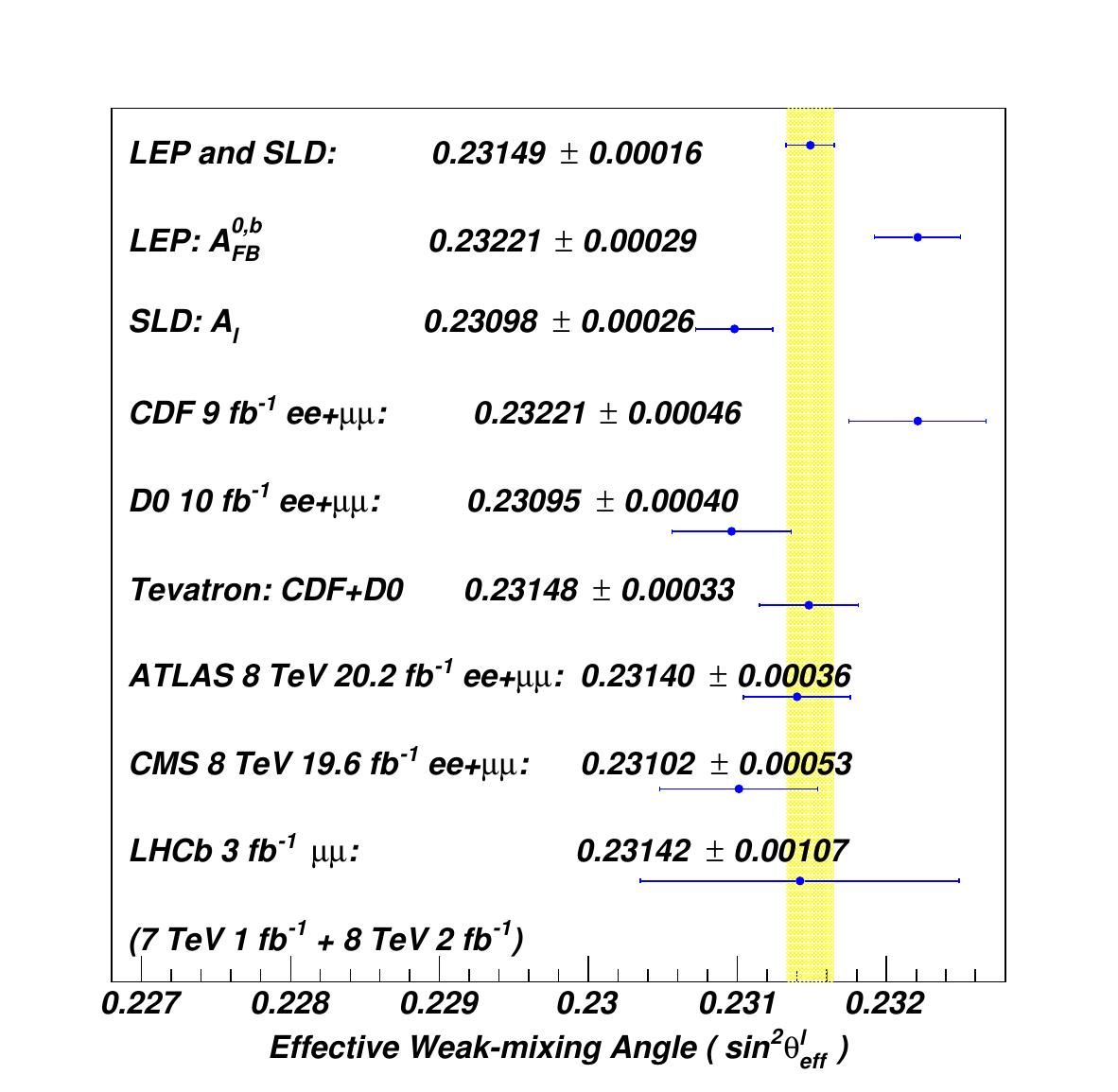}
\caption{Previously measured $\effstw$ from LEP, SLD, Tevatron and LHC. }
\label{fig:stwHistory}
\end{center}
\end{figure}

$\effstw$ has been measured in the past two decades using the $f_i\bar{f_i}\rightarrow Z/\gamma^* \rightarrow f_j\bar{f_j}$ 
productions. The results are 
shown in Figure~\ref{fig:stwHistory}. The most precise determinations of $\effstw$ come from the electron-positron 
colliders, which are 0.23221$\pm$0.00029 from the combined LEP $b$ quark production, and 0.23098$\pm$0.00026 
from the SLD~\cite{LEP-SLD}. After that, a similar precision was achieved at the proton-antiproton collider Tevatron, giving 
0.23148$\pm$0.00033~\cite{Tevatron-combine}. $\effstw$ are also measured by 
ATLAS, CMS and LHCb collaborations~\cite{ATLASstw, CMSstw, LHCbstw}. 
$\effstw$ was extracted from the forward-backward charge 
asymmetry ($A_{FB}$, to be introduced in the following sections) 
in all those measurements above, except for the SLD result which also used the left-right 
polarization asymmetry. 

Previous measurements achieved an relative precision at $\mathcal{O}(0.1\%)$ with respect to the central 
value of $\effstw$. It played an important role in the global fitting of the SM electroweak sector in the past years. However, 
the experimental precision, which is generally limited by the size of the data sample,  
is much worse than the precision of the theoretical calculations. At two-loop level, the 
uncertainty in $\effstw$ calculation is reduced to 0.00004~\cite{PDGstw}. It would be essential to improve the 
experimental precision on $\effstw$ to be comparable to the theoretical calculations. 

Though a large data sample will be collected at the LHC in the next 10 years, it will be very difficult to have the 
uncertainty on $\effstw$ to be smaller than 0.00010 using the LHC data. 
At hadron colliders, the initial state fermions of the neutral-current 
interactions are quarks and antiquarks acting as partons in the hadrons. 
Their effective momentum are described by the parton distribution functions (PDFs), which extrapolate large uncertainties 
to the $\effstw$ extraction. In the recent LHC measurements, the PDF-induced uncertainties on $\effstw$ 
are larger than $0.00020$~\cite{ATLASstw, CMSstw}, and will become the most leading uncertainty in the future. 
Such uncertainty will not be naturally reduced as the LHC data is introduced in the PDF global analysis, due to 
a strong correlation between the PDF and the $\effstw$ in the LHC observations~\cite{CPCstudy}. The QCD-induced 
uncertainty is also larger than $0.00010$ at the LHC~\cite{CMSstw}, extrapolated via soft-gluon radiations in the 
initial state. In conclusion, it would be most likely to achieve a high precision determination on $\effstw$ at 
next generation electron-positron colliders, which are generally free from PDF and QCD, and have the 
capability to generate a large data sample. 

In this paper, we study the measurement of $\effstw$ at the proposed Circular Electron Positron Collider (CEPC). 
CEPC is a powerful machine providing physics interactions with high energy electron-positron initial state~\cite{CEPCCDR}. 
It is proposed to have a 2-year running plan around the $Z$ boson mass pole.
We focus on the precision of $\effstw$ in lepton and $b$ quark final states, with both statistical uncertainty and 
potential experimental systematics taken into account. 
~\\

\section{Forward-backward charge asymmetry}

\begin{figure}[!hbt]
\begin{center}
\epsfig{scale=0.4, file=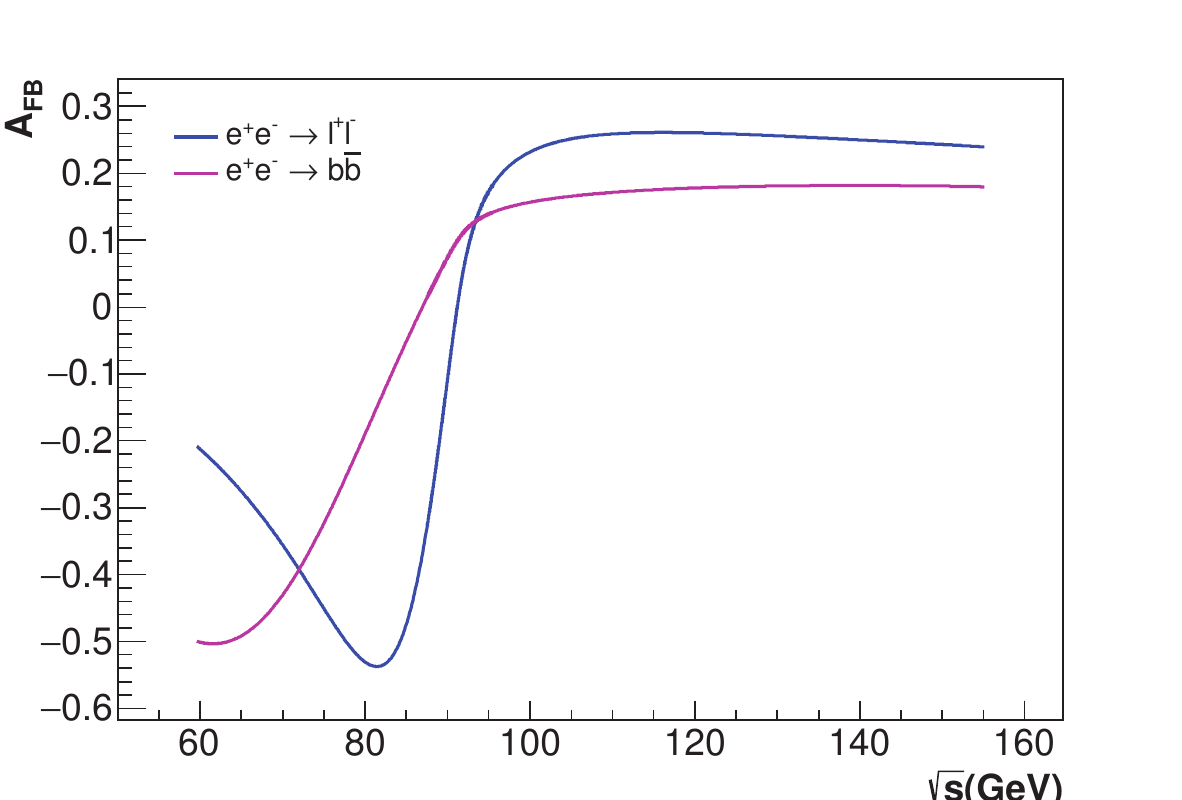}
\caption{The $A_{FB}$ spectrum as a function of $\sqrt{s}$ for $e^+e^-\rightarrow Z/\gamma^* \rightarrow \ell^+\ell^-$ and 
 $e^+e^-\rightarrow Z/\gamma^* \rightarrow b\bar{b}$ productions.}
\label{fig:AFBvsM}
\end{center}
\end{figure}

\begin{figure}[!hbt]
\begin{center}
\epsfig{scale=0.4, file=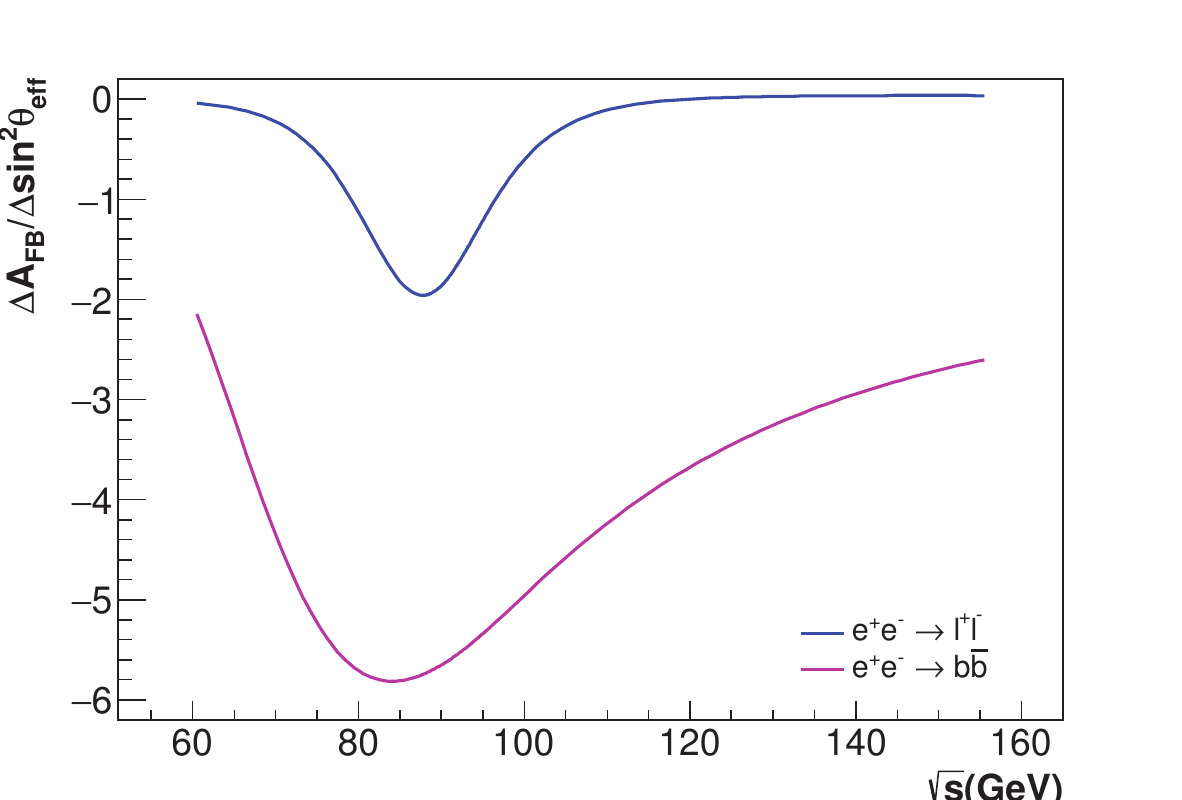}
\caption{The sensitivity of $S = \partial A_{FB} / \partial \effstw $ as a function of $\sqrt{s}$ 
for $e^+e^-\rightarrow Z/\gamma^* \rightarrow \ell^+\ell^-$ and  
 $e^+e^-\rightarrow Z/\gamma^* \rightarrow b\bar{b}$ productions.}
\label{fig:sensitivity}
\end{center}
\end{figure}

The forward-backward charge asymmetry $A_{FB}$ of the 
$f_i\bar{f_i} \rightarrow Z/\gamma^* \rightarrow f_j\bar{f_j}$
process is an ideal experimental observable
to probe the electroweak interaction with a high precision.
It is defined as
\begin{eqnarray}
 A_{FB} = \frac{N_F - N_B}{N_F + N_B},
\end{eqnarray}
where $N_F$ and $N_B$ are the numbers of the forward and backward events, judged by the 
scattering angle $\theta_{ij}$ formed by the directions of the initial state negative charged electron beam and the final state fermion.
Events with $\cos\theta_{ij}>0$ are classified as forward ($F$) and those with $\cos\theta_{ij}<0$ as 
backward ($B$). 
At the CEPC, the initial state fermions are electrons and positrons, while the final state fermions can be 
leptons and quarks.
The asymmetry arises from the interference between vector and axial vector coupling terms, and 
precisely governed by $\effstw$. The value of $A_{FB}$ changes with the center-of-mass energy $\sqrt{s}$.
Figure~\ref{fig:AFBvsM} shows the $A_{FB}$ spectrum as a function of $\sqrt{s}$ for  
different productions. 
The predictions are calculated using the effective born approximation package {\sc zfitter} corresponding to 
next-to-next-to-leading order (NNLO)
radiative corrections~\cite{ZFITTER}. $A_{FB}$ is very sensitive to $\effstw$ around the $Z$ mass pole. 
In the low mass region, the asymmetry is approaching zero as $\sqrt{s}$ 
goes lower due to the rising contribution of the photon exchange. In the high mass region, the asymmetry 
has roughly a constant value dominated by the interference between the $\gamma$ and $Z$ boson exchanges. 
Therefore, the sensitivity of off-pole $A_{FB}$ to $\effstw$ is significantly reduced. The sensitivity, defined as 
$$S = \partial A_{FB} / \partial \effstw,$$ 
is given as a function of $\sqrt{s}$ in Figure~\ref{fig:sensitivity} for $b$ quark and 
lepton productions as an example. Predictions are calculated using {\sc zfitter} as well. 
In the following sections, we estimate the uncertainty on $\effstw$ based on 
the sensitivity $S$ of $A_{FB}$ to $\effstw$.

~\\

\section{Experimental observation on $A_{FB}$ and extraction of $\effstw$}

The observed asymmetry, denoted as $A^\text{obs}_{FB}$, could be biased due to the imperfect 
detector performance. Three major contributions of the potential experimental 
systematics are discussed in this work, which come from the $\sqrt{s}$ determination, the charge measurement and the inefficiency of particle 
reconstruction and selection. These systematics are generally small.
At lepton colliders, $\sqrt{s}$ of an event can be precisely controlled by the beam energy, instead of reconstructed 
from the final state particles measured in the detector. According to the 
CEPC conceptual design report, the uncertainty of the electron and positron beam energy can be 
controlled around 100 KeV~\cite{CEPCAcc}, extrapolating a relative uncertainty on $\effstw$ 
much smaller than $0.01\%$. The inefficiency of particle reconstruction and selection could have a large 
effect especially in quark productions. However,  $A_{FB}$ is defined as a relative asymmetry where 
total cross section perfectly cancelled. Therefore, the limited efficiency only enlarges the statistical uncertainty 
of the $A_{FB}$ observation and $\effstw$ extraction and causes no systematics, as long as there is 
no difference between the forward and backward event efficiencies.
With CEPC's large data sample, the statistical uncertainty will be negligibal anyway.

A more complicated case is 
the charge mis-identification of the final state particles, which could cause both systematic extrapolation and 
statistical uncertainty increase.
The forward and backward categories are classified according to the charge of the final state particles.
If an event has a probability of $f$ to be wrongly classified 
between forward and backward due to the mis-identification of charge, the observed $A_{FB}$ will be 
diluted from the original $A_{FB}$, written as:
\begin{eqnarray}\label{eq:AFBdef}
 A^\text{mis-q}_{FB} = (1-2f) A_{FB}.
\end{eqnarray}
When $f=50\%$, there will be no observed asymmetry. Such dilution causes reduction in the $A_{FB}$ to 
$\effstw$ sensitivity, appearing as an enlarged statistical uncertainty. 
For the selected $e^+e^- \rightarrow Z/\gamma^* \rightarrow f\bar{f}$ events, 
$f$ can be determined from 
the following relationship
\begin{eqnarray}
 N_{ss} = 2\omega(1-\omega)\cdot N_\text{total},
\end{eqnarray}
 where $N_{ss}$ is the number of selected events with same charge sign of the final state fermions, 
 while $N_\text{total}$ is the total number of 
 selected events. $\omega$ is the probability to mis-identify the charge of a single fermion, and we have $f = \omega^2/(\omega^2+(1-\omega)^2)$.
 The precision of the $f$ determination is dominated by the statistics. Considering 
 the large data sample at the CEPC, $f$ could be precisely determined by this data-driven method 
 and causes very small systematics. Besides, the final state 
 fermions are usually required to have opposite charge in order to suppress the mis-identification of 
 the forward and backward categories. 

According to the definition of $A_{FB}$ in Eq.~\ref{eq:AFBdef} and assuming that the 
value of $A_{FB}$ around $Z$ pole is close to zero, the statistical uncertainty on the observed 
asymmetry $A^\text{obs}_{FB}$ is approximately written as
\begin{eqnarray}
 \delta A^\text{obs}_{FB} = \sqrt{\frac{1-\left(A^\text{obs}_{FB}\right)^2}{N}} \approx \sqrt{\frac{1}{N}} ,
\end{eqnarray}
where $N$ is the number of the selected events. 
Taking the above effects into consideration, the statistical uncertainty of $\effstw$ measured from $A_{FB}$
can be expressed as:
\begin{eqnarray}\label{eq:statUnc}
\delta \effstw = \sqrt{\frac{1}{N}} \cdot \sqrt{\frac{1}{\epsilon\cdot (1-2f)^2}} \cdot \frac{1}{|S|} ,
\end{eqnarray}
where $\epsilon$ is the overall efficiency of detecting and selecting an $e^+e^-\rightarrow Z/\gamma^* \rightarrow \ell^+\ell^-$ event. 
The term $\epsilon\cdot (1-2f)^2$ is defined as the tagging power parameter. For lepton final states, the overall efficiency 
is very close to $100\%$, and $f$ is negligibly small. Therefore the tagging power is almost $100\%$~\cite{CEPCCDR}. 
For $b$ quark final state, 
it is much more complicated. It is difficult to determine the $b$ quark charge by measuring the final state jet. 
To have a better charge measurement, only a small part of the $b$ quark production events, where 
$b$ quarks decay to leptons or Kaons, could be used in the $\effstw$ measurement. Therefore, the tagging power parameter 
for $b$ quark productions needs to be optimized between the overall efficiency and the charge mis-identification probability. 
According to the CEPC simulation study~\cite{CEPCCDR,CEPCtagging}, with a selection of $b$ quark with $98\%$ purity, the optimized tagging power for $b$ quark production is 
0.088.

The number of the selected events $N$ depends on the luminosity of the proposed running plan, and the cross section of each 
channel of the $Z$ boson decay. The latest CEPC studies give a proposal of 2-year running period around the $Z$ boson 
mass pole, with 50 ab$^{-1}$ integrated luminosity per year. In another word, CEPC can provides $1.7\times 10^{11}$ $Z$ boson events every month. Considering 
the branching ratio~\cite{PDGstw} and Eq.~\ref{eq:statUnc}, the expected statistical uncertainty 
of $\effstw$ measured from lepton final state ($ee$+$\mu\mu$) 
is $\delta \effstw(\ell) = 5 \times 10^{-6}$ using 1 month data. For $b$ quark productions, the uncertainty is 
$\delta \effstw(b) = 4 \times 10^{-6}$ using 1 month data. 
This uncertainty can be further reduced using the proposed 2-year data sample. However, it would be more useful to 
run at different collision energy point off-pole rather than simply collecting data at the very peak of the $Z$ mass line shape. 
When changing the collision energy in this study, the cross section of the $Z$ boson production 
is altered according to its mass line shape. The drop of the instantaneous luminosity is estimated 
approximately as the third power of the increase of the collision energy~\cite{CEPCAcc}. The expected statistical 
uncertainties of $\effstw$ with 1 month data collection at different collision energy points are 
summarized in Table~\ref{tab:results}
\linespread{\LINESPREADSIZE}
\begin{table}[H]
  
\begin{tabular}{ccc}
\topL
    collision energy & $\delta\effstw$ in &  $\delta\effstw$ in \\
    (GeV) & lepton final state & $b$ quark final state \\ 
\midL
70 & $1.5\times 10^{-4}$ & $4.1\times 10^{-5}$ \\
75 & $6.8\times 10^{-5}$& $3.3\times 10^{-5}$\\
92 & $4.9\times 10^{-6}$&  $3.5\times 10^{-6}$ \\
105 & $1.7\times 10^{-4}$& $2.7\times 10^{-5}$ \\
115 & $2.0\times 10^{-3}$& $4.8\times 10^{-5}$ \\
130 & $4.0\times 10^{-3}$& $9.8\times 10^{-5}$ \\
\botL
\end{tabular}
\caption{\small The expected statistical uncertainties on $\effstw$. Results are estimated according to 1 month data 
collection.}
\label{tab:results}

\end{table}
As we can see, both lepton final state and $b$ quark final state can provide precise determination on $\effstw$ at the 
$Z$ boson mass pole. For the measurement of the energy running effect, $b$ quark production has higher precision 
because the sensitivity $S$ off-pole drops much slower than the lepton final state cases. To make a conservative 
estimation, 
the precision on the $\sin^2\theta^\ell_\text{eff}$ determination, considering both the statistical uncertainty
and the experimental systematics at the CEPC, can be 0.00001 in both lepton and $b$ quark productions with 1 month data collection.
The $\effstw$ can be measured as a function of collision energy up to 130 GeV, with a precision of around $0.0001$ from the 
$b$ quark productions.

Due to the contribution of the $t$-channel and the $s$-$t$ interference in the electron final state,
the uncertainty of the theoretical calculation in the electron final state can be very large, giving $0.00085$ on the weak mixing angle according to Ref.~\cite{LEP-SLD}.
However, such uncertainty only affects the dielectron events.
For other channels, the residual theoretical uncertainties can be 
much smaller, giving $0.00006$ on the weak mixing angle \cite{LEP-SLD}.
Therefore, the best precision of the $\sin^2\theta^\ell_\text{eff}$ determination relies on the muon channel.
The uncertainty on the calculations of the $e^+e^-\rightarrow f\bar{f}$ process is much larger
than the statistical uncertainty and the experimental systmatics,
thus will be the major source limiting the final precision of the experimental measurement of $\sin^2\theta^\ell_\text{eff}$.
However, it will still be much better than the expected precision at the hadron colliders.

\section{Supplementary discussion on the $\effstw$ measurement}

Aside from electon, muon and b quark channel $A_{FB}$ measurement, other
channels such as c quark can also be utilized to extract $\effstw$.
With the predicted sensitivity $S_{f\bar{f}}$ listed in table \ref{tab:results-Sensitivity},
one can estimate the precision of $\effstw$ from $A_{FB}$ measurement using
Eq. \ref{eq:statUnc}, after the performance research for different
final state particles. For instance, a recent CEPC simulation study~\cite{CEPCtagging_new}
used leading particle and weighted jet charge combined information
to give a higher performance of heavy flavor jet charge measurement,
and the bare tagging power\footnote{Here ``bare'' means the tagging power was estimated using a pure
b/c jet, while the flavor tagging of b/c filnal state in Zbb/Zcc event
was not taken into consideration.} of b/c quark was determined.
The tagging power of b quark final state is doubled, 
therefore the estimated $\delta\effstw$ with b quark final state 
is $2.5\times10^{-6}$
(with 1 month data collection at Z pole). However, for the c quark
circumstance, due to the low purity of c flavor tagging, the estimation
will need detailed simulation study of flavor tagging using $Z\rightarrow q\bar{q}$
samples.
\linespread{\LINESPREADSIZE}
\begin{table}[H]
  \centering{}%
  \begin{tabular}{ccccccc}
  \topL
  $\sqrt{s}$ &  $S$ of  & $ S$ of  & $ S$ of  &  $S$ of  &  $S$ of  &  $S$ of \\
  (GeV) &  $A_{FB}^{e/\mu}$ &  $A_{FB}^{d}$ &  $A_{FB}^{u}$ &  $A_{FB}^{s}$ &  $A_{FB}^{c}$ &  $A_{FB}^{b}$\\
  \midL
  70 &  0.224 &  4.396 &  1.435 &  4.403 &  1.445 &  4.352\\
  75 &  0.530 &  5.264 &  2.598 &  5.269 &  2.616 &  5.237\\
  92 &  1.644 &  5.553 &  4.200 &  5.553 &  4.201 &  5.549\\
  105 &  0.269 &  4.597 &  1.993 &  4.598 &  1.994 &  4.586\\
  115 &  0.035 &  3.956 &  1.091 &  3.958 &  1.087 &  3.942\\
  130 &  0.027 &  3.279 &  0.531 &  3.280 &  0.520 &  3.261\\
  \botL
  \end{tabular}
  \caption{Sensitivity $S$ of different final state particles.}
  \label{tab:results-Sensitivity}
  \end{table}
  
  Tau lepton is the only final state fermion we've known whose polarizarion($P_{\tau}$)
  can be measured at an unpolarized leptonic collider~\cite{LEP-SLD} with
  \begin{eqnarray*}
    P_{\tau}=\frac{\mathrm{d}(\sigma_r-\sigma_l)}{\mathrm{d}\cos{\theta}}\Big/
    \frac{\mathrm{d}(\sigma_r+\sigma_l)}{\mathrm{d}\cos{\theta}},
    \label{eq:PTauMeasurement}
    \end{eqnarray*}
  where $\sigma_{r/l}$ is the cross section for producing right/left-handed
  final state tau leptons.
  And $P_\tau$ is connected to $\effstw$ by
  \begin{eqnarray}
  P_{\tau}=-\frac{\mathcal{A}_\tau\cdot(1+\cos^2{\theta})+\mathcal{A}_e\cdot(2\cos{\theta})}
  {(1+\cos^2{\theta})+\mathcal{A}_\tau\mathcal{A}_e\cdot(2\cos{\theta})},
  \label{eq:PTauTheory}
  \end{eqnarray}
  where
  \begin{eqnarray*}
    \mathcal{A}_f=\frac{2g_V^fg_A^f}{(g_V^f)^2+(g_A^f)^2}=\frac{2g_V^f/g_A^f}{1+(g_V^f/g_A^f)^2}
  \end{eqnarray*}
  is the asymmetry parameter.
  This property
  was utilized by LEP to perform rather independent measurement for
  $\effstw$. Compared to the whole lepton channel, the statistics of tau
  channel is small, and the efficiency and purity of tau reconstruction
  are low. But with a very high sensitivity of $P_{\tau}$ to $\tau-Z$
  vector coupling constant, it can give a high precision extraction
  of $\effstw$.
  
\begin{figure}[!hbt]
    \begin{center}
    \epsfig{scale=0.35, file=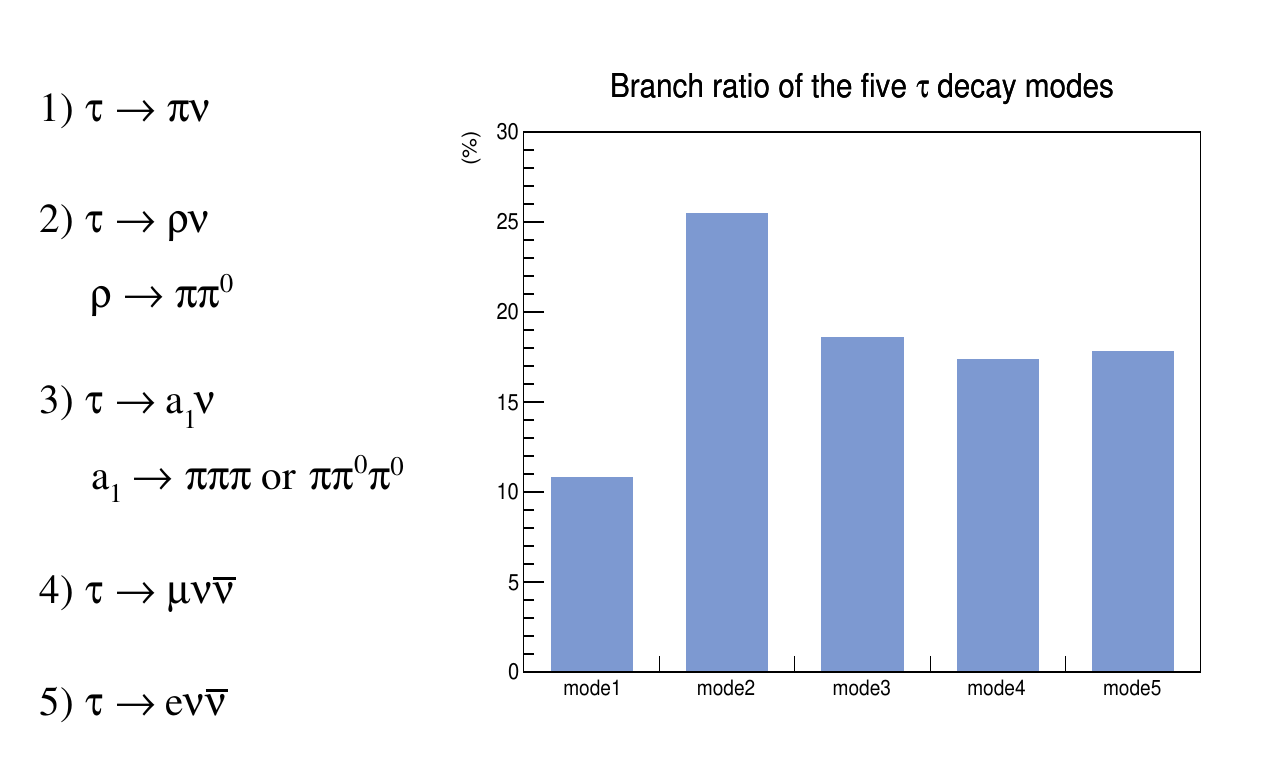}
    \caption{The relative contributions of different decay modes in the $\tau$ final state.
    }
    \label{fig:tau_decay_mode}
    \end{center}
  \end{figure}

\begin{figure}[!hbt]
    \begin{center}
    \epsfig{scale=0.23 , file=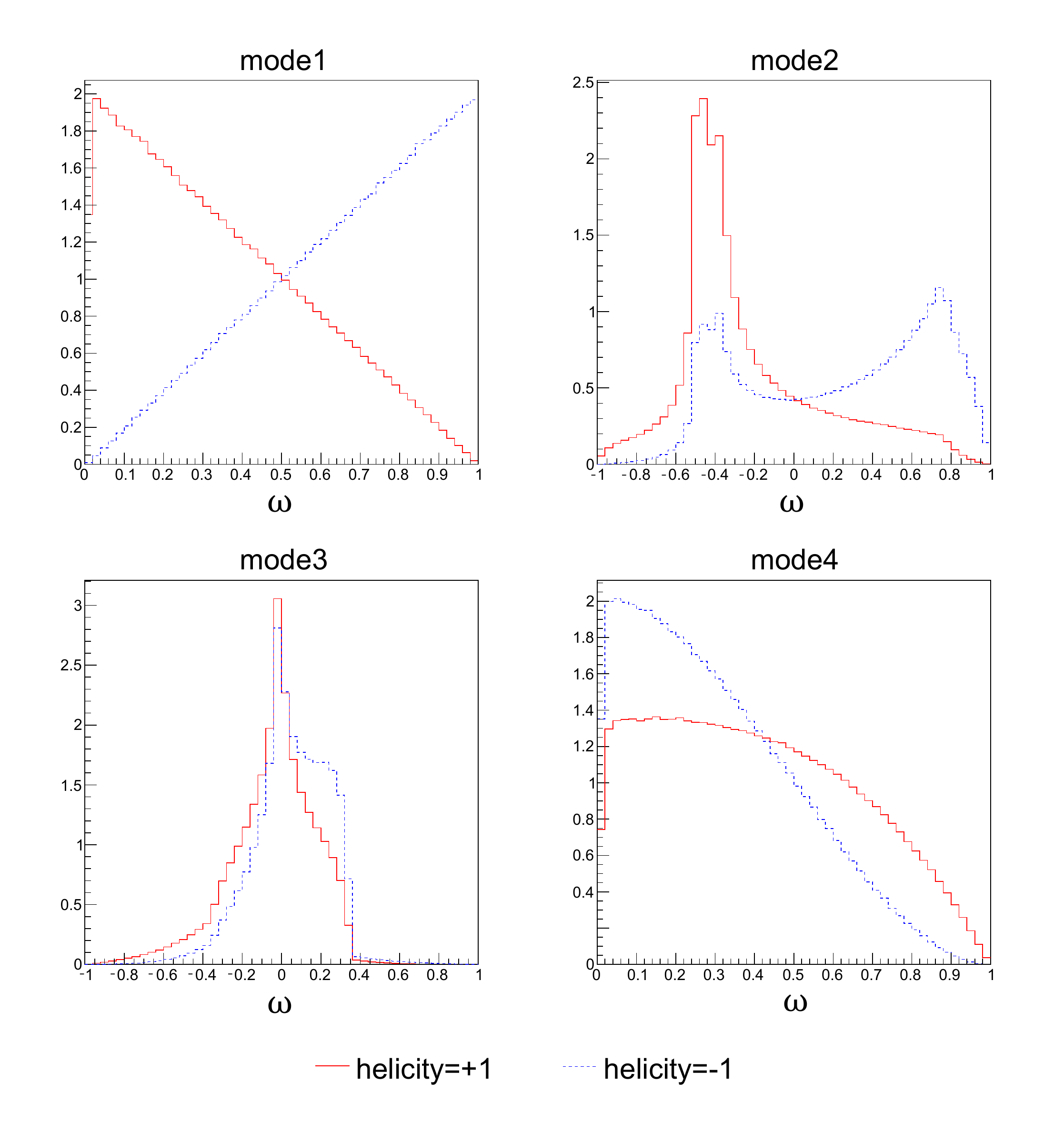}
    \caption{Kinematic spectrum of different tau decay modes.
    The red solid line and blue dashed line represent the kinematic spectrum of 
    taus with $helicity=+1$ and $-1$, respectively.
    All the spectra are genarated using {\sc{pythia8}} genarator 
    and {\sc{tauola}} interface.}
    \label{fig:tau_each_mode}
    \end{center}
  \end{figure}

  Measurement of $P_{\tau}$ is based on the fact that $\tau$ has a
  short lifetime and that the kinematic spectrum of its decay production
  is different when tau has different helicity (shown in Figure \ref{fig:tau_decay_mode} and
  Figure \ref{fig:tau_each_mode}).
  We use {\sc pythia8} program~\cite{PYTHIA} to generate $e^{+}e^{-}\rightarrow\tau^{+}\tau^{-}$
  events, and then use {\sc tauola} interface~\cite{TAUOLA} to decay the taus.
  
  Using two templates with $helicity=+1$ and $-1$, respectively, we can perform
  fit to the pseudo-data, whose $P_\tau$ is a given number.
  Due to limited computing resources, only $2\times10^8$ $Z\rightarrow\tau\bar{\tau}$
  events are generated for pseudo-data and each template.
  The extrapolation of the fitting results shows that the statistical
  uncertainty on $\effstw$ with 1 month data collection
  at Z pole is $2.15\times10^{-6}$ using $P_\tau$ measurement.

Experimentally, the $\tau$ lepton is reconstructed from its daughter particles in the decay,
thus relies on the precision of the measurement of the particle energy and the background control.
It makes the detector systematics extrapolate more significantly to the weak mixing angle than in other channels,
According to the stydy of LEP~\cite{LEP-SLD}, the systematics in the $\tau$ measurement
is at an order of $\mathcal{O}(10^{-4})$ on the weak mixing angle.
Given that the statistical uncertainty at the CEPC would be much smaller, the total uncertaintt in the $\tau$ channel measurement will be dominated by the systematics.

\section{Conclusion}
We present an estimation of the precision on $\effstw$ determination at the CEPC in the lepton final state and $b$ quark 
final state.
With a high instantaneous luminosity, the statistical uncertainty can be reduced to be negligible.
The experimental sysmtematics are also negligible in general since $A_{FB}$ is defined as a relative
asymmetry so that the systematics cancel out. The dominant uncertainty will come from the theoretical calculation on the $e^-e^+\rightarrow f\bar{f}$ process.
As a result, the precision of $\sin^2\theta^\ell_\text{eff}$ can be improved to $\mathcal{O}(10^{-5})$,
from the current precision of $\mathcal{O}(10^{-4})$ at the LEP, SLC and Tevatron.
Due to a large model uncertainty from QCD calculations and PDF modeling, such precision is difficult to be achieved 
using the LHC data in the future.
This precision will, for the first time, be comparable to the precision of the theoretical calculation of $\sin^2\theta^\ell_\text{eff}$
with radiative corrections at two-loop level, meaning that the precision of the SM electroweak global fit can be significantly improved.
One thing worth emphasizing is that the high precision measurement of $\effstw$ at the 
CEPC is essential to the QCD studies at the LHC. As discussed in the introduction section, the observation of proton structure 
and electroweak symmetry breaking is highly correlated in $pp(q\bar{q})\rightarrow Z/\gamma^* \rightarrow \ell^+\ell^-$ events. 
In Ref.~\cite{CPCstudy}, it is proved that the single $Z$ boson production can provide unique information of the relative difference 
between quarks and antiquarks. However, it is not available yet in the PDF global fitting due to a large uncertainty induced 
by the experimental determination of $\effstw$. Using the CEPC's electron-positron interaction, the measured $\effstw$ can 
be used as high precision input in the PDF global fitting, fixing the electroweak calculations in predicting 
the single $Z$ boson production cross sections. 

One last thing to be emphasized is that our analysis uses high purity $b\bar{b}$ sample
to exclude the contamination of other quark flavors. With a properly designed working point
for jet flavor tagging, we can also use other quark flavors to measure $\effstw$.
Future development of detector optimization and advanced reconstruction algorithm,
especially those based on machine learning, for example those has been used recently
at CMS experiment~\cite{CMS-particle-net}, could also boost the relevant performances.

~\\

\section{Acknowledgements}

We thank Dr. Zhijun Liang from the Institution of High Energy Physics Chinese Academy of Science, for the helpful discussions.
This work was supported by the ``USTC Research Funds of the Double First-Class Initiative'', 
the International Partnership Program of Chinese Academy of Sciences 
(Grant No. 113111KYSB20190030), and the Innovative Scientific Program of Institute of High Energy Physics.
~\\

\section*{Appendix}

\subsection{Cross section of $e^{+}e^{-}\rightarrow Z\rightarrow f\bar{f}$}

\linespread{\LINESPREADSIZE}
\begin{table}[hbt]
\centering{}%
\begin{tabular}{ccccccc}
\topL
$\sqrt{s}(GeV)$ &$\sigma_{\mu}(mb)$ &$\sigma_{d}(mb)$ &$\sigma_{u}(mb)$ &$\sigma_{s}(mb)$ &$\sigma_{c}(mb)$ &$\sigma_{b}(mb)$\\
\midL
70 &0.039 &0.032 &0.066 &0.031 &0.058 &0.028\\
75 &0.039 &0.047 &0.073 &0.046 &0.065 &0.043\\
92 &1.196 &5.366 &4.228 &5.366 &4.222 &5.268\\
105 &0.075 &0.271 &0.231 &0.271 &0.227 &0.265\\
115 &0.042 &0.135 &0.122 &0.135 &0.118 &0.132\\
130 &0.026 &0.071 &0.068 &0.071 &0.066 &0.069\\
\botL
\end{tabular}\caption{
  Cross section of process $e^{+}e^{-}\rightarrow f\bar{f}$ calculated using the {\sc zfitter}
  package. Values of the fundamental parameters are set as 
  $m_{Z}=91.1875\mathrm{GeV}$, $m_{t}=173.2\mathrm{GeV}$,
  $m_{H}=125\mathrm{GeV}$, $\alpha_{s}=0.118$ and $m_{W}=80.38\mathrm{GeV}$.}
\label{tab:xsection}
\end{table}

\end{document}